\documentclass[twocolumn,apl,amsmath,jmp,amssymb,reprint,floatfix,superscriptaddress]{revtex4}
\usepackage{appendix}
\usepackage{graphicx}
	\usepackage{amssymb}
	\usepackage{amsmath}
	\usepackage{epstopdf}
	\usepackage[plain]{fancyref}
	\usepackage[colorlinks = true]{hyperref}
	\usepackage[section,subsection,subsubsection]{extraplaceins}

\begin{document}

\title{Direct Imaging of Dynamic Glassy Behavior in a Strained Manganite Film}
\author{Worasom Kundhikanjana}
	\affiliation{Department of Applied Physics and Geballe Laboratory for Advanced Materials, Stanford University, Stanford, California 94305, USA}
	\affiliation{School of Physics, Institute of Science, Suranaree University of Technology, Nakron Ratchasima, Thailand}
\author{Zhigao Sheng}
	\affiliation{RIKEN Center for Emergent Matter Science (CEMS), Wako 251-0198, Japan}
	\affiliation{High Magnetic Field Laboratory of Chinese Academy of Science, 
	Hefei 230031, P.R. China and Collaborative Innovation Center of Advanced Microstructures, Nanjing 210093, P.R. China }
\author{Yongliang Yang}
	\affiliation{Department of Applied Physics and Geballe Laboratory for Advanced Materials, Stanford University, Stanford, California 94305, USA}
\author{Keji Lai}
	\affiliation{Department of Physics, University of Texas at Austin, Austin, Texas 78712, USA}
\author{Eric Yue Ma}
	\affiliation{Department of Applied Physics and Geballe Laboratory for Advanced Materials, Stanford University, Stanford, California 94305, USA}
\author{Yong-Tao Cui}
	\affiliation{Department of Applied Physics and Geballe Laboratory for Advanced Materials, Stanford University, Stanford, California 94305, USA}
\author{Michael A. Kelly}
	\affiliation{Department of Applied Physics and Geballe Laboratory for Advanced Materials, Stanford University, Stanford, California 94305, USA}
\author{Masao Nakamura}
	\affiliation{RIKEN Center for Emergent Matter Science (CEMS), Wako 251-0198, Japan}
\author{Masashi Kawasaki}
	\affiliation{RIKEN Center for Emergent Matter Science (CEMS), Wako 251-0198, Japan}
	\affiliation{Department of Applied Physics and Quantum Phase Electronics Research Center (QPEC), University of Tokyo, Tokyo 113-8656, Japan.}
\author{Yoshinori Tokura3}
	\affiliation{RIKEN Center for Emergent Matter Science (CEMS), Wako 251-0198, Japan}
	\affiliation{Department of Applied Physics and Quantum Phase Electronics Research Center (QPEC), University of Tokyo, Tokyo 113-8656, Japan.}

\author{Qiaochu Tang}
	\affiliation{State Key Lab of Transducer Technology, Shanghai Institute of Microsystem and Information Technology, Chinese Academy of Sciences, Shanghai 200050, China}
\author{Kun Zhang}
	\affiliation{State Key Lab of Transducer Technology, Shanghai Institute of Microsystem and Information Technology, Chinese Academy of Sciences, Shanghai 200050, China}
\author{Xinxin Li}
	\affiliation{State Key Lab of Transducer Technology, Shanghai Institute of Microsystem and Information Technology, Chinese Academy of Sciences, Shanghai 200050, China}
\author{Zhi-Xun Shen}
	\affiliation{Department of Applied Physics and Geballe Laboratory for Advanced Materials, Stanford University, Stanford, California 94305, USA}
\date{\today}


\begin{abstract}
Complex many-body interaction in perovskite manganites gives rise to a strong competition between ferromagnetic metallic and charge ordered phases with nanoscale electronic inhomogeneity and glassy behaviors. Investigating this glassy state requires high resolution imaging techniques with sufficient sensitivity and stability. Here, we present the results of a near-field microwave microscope imaging on the strain driven glassy state in a manganite film. The high contrast between the two electrically distinct phases allows direct visualization of the phase separation. The low temperature microscopic configurations differ upon cooling with different thermal histories. At sufficiently high temperatures, we observe switching between the two phases in either direction. The dynamic switching, however, stops below the glass transition temperature. Compared with the magnetization data, the phase separation was microscopically frozen, while spin relaxation was found in a short period of time.  
	\end{abstract}

\maketitle 

A glass is formed by rapid cooling of a viscous liquid, resulting in a supercooled liquid with no crystallinity\cite{c1}. Generally, such supercooled state can occur in many systems with first-order phase transition. In these systems, there are multiple competing states separated by a thermal barrier near the transition temperature. By rapid cooling, the system can be trapped in the non-favorable state, resulting in slow relaxation and cooling rate dependent behaviors. Furthermore, the existence of complex energy landscapes often leads to different low-temperature states even under the same cooling process, known as non-ergodicity. Perovskite manganites are a good example of systems with such dynamics. In half-doped manganites, the transition from the charge-ordered insulating state (CO-I) to the ferromagnetic metallic state (FM-M) is first-order in nature, while the energetic proximity between the two crystalline states often results a phase-separation\cite{c2,c3} with one of states being metastable. The metastability gives rise to relaxation behaviors \cite{c4,c5} and dependences on cooling histories \cite{c6,c7,c8}. Aspects of spin-glass-like behaviors are also found, such as frequency dependent AC susceptibility \cite{c9}. The phase-separated (PS) state is also highly susceptible to local parameters such as strain \cite{c10} and disorder \cite{c9}. 

\begin{figure}[t]
	\centering
		\includegraphics[width=3.2in]{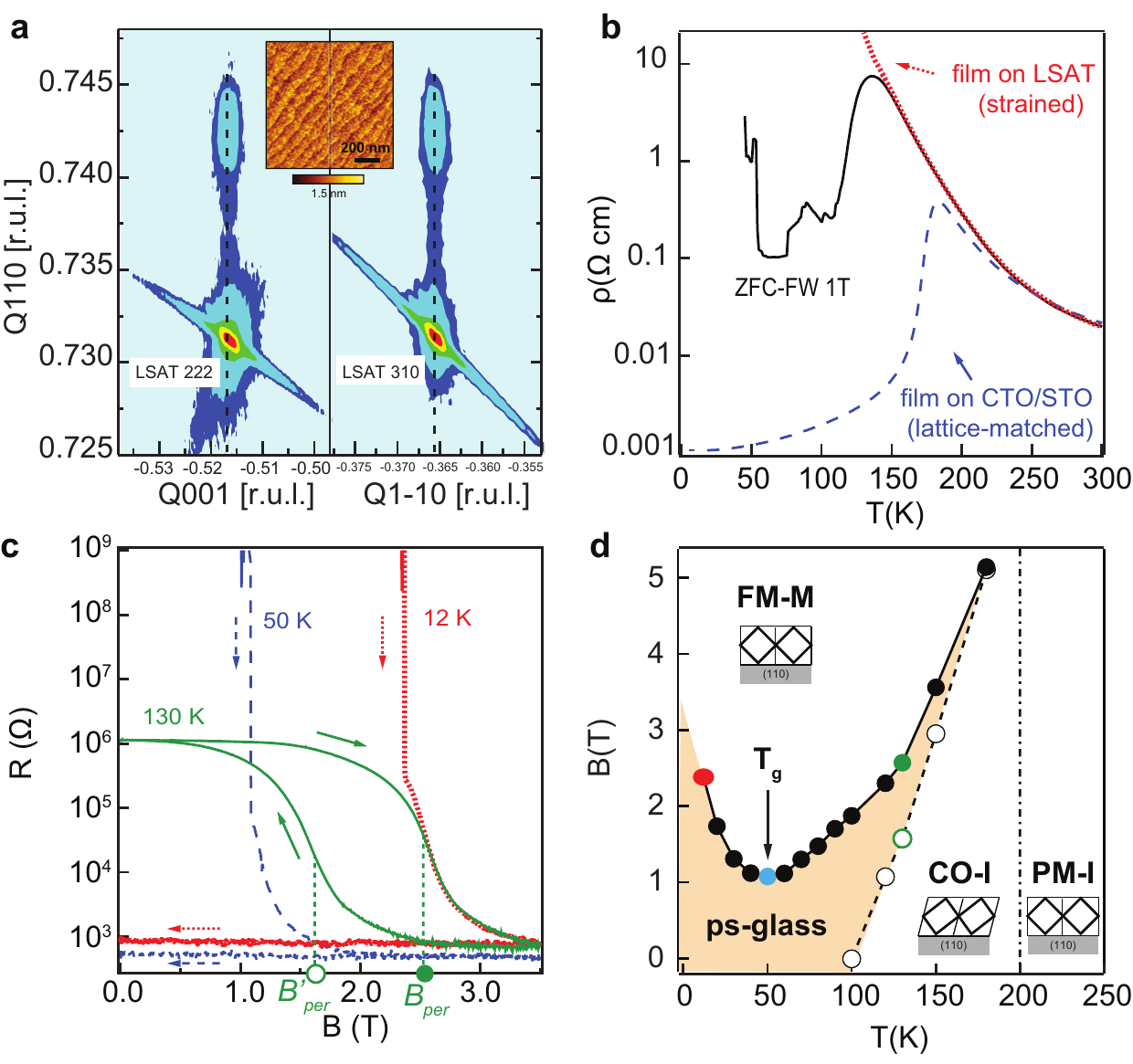}
		\caption{(a) Reciprocal space map of the PCSMO film grown on LSAT. (b) Resistivity of the strained PCSMO/LSAT film (dotted-red) and a relaxed film on the CTO/STO substrate (dash-blue). The solid-black line was taken during field warming after ZFC with $B$ = 1 T. (c) Magnetoresistance after ZFC to 12 K, 50 K and 130 K. The full and empty circles indicate $B_{per}/B?_{per}$ for the forward and backward sweeps, respectively. (d) B-T phase diagram constructed from$B_{per}$ and $B?_{per}$ at various temperature.  The insets show schematic views of the film on LSAT (110) in each phase. Both PM-I and FM-M phases have cubic structures, while the CO-I phase is distorted orthorhombic.}
			\label{fig1}
\end{figure}

Many models for manganite glass were constructed using macroscopic measurements\cite{c11}. However, the PS state is necessary for understanding the physics of manganites \cite{c2}. Transport measurements, therefore, have clear drawbacks. For example, macroscopic magnetization cannot distinguish intra domain relaxation from phase switching \cite{c12}. Although the dynamics of PS state may be inferred from the transports through a manganite nanowire \cite{c13}, but these measurements are limited to a narrow temperature range with sufficient conductivity. To date, the PS state with distinct physical properties has been visualized by various microscopy techniques such as scanning tunneling microscopy \cite{c14}, electron microscopy \cite{c15}, photoemission spectroscopy \cite{c16}, magnetic force microscopy \cite{c17,c18}, and recently near-field microwave impedance microscopy (MIM) \cite{c19}. However, most of these studies focus on the static phase-separation. Direct visualization of the dynamic behaviors in the PS manganites is yet to be addressed.

In this paper, we present the results of a MIM study on a  Pr$_{0.55}$(Ca$_{0.75}$Sr$_{0.25}$ (PCSMO) film. The high contrast between the two electrically distinct phases allows detailed study of the PS states. The large field of view and the high measurement stability enable tracking of the dynamic behavior of several FM-M domains in a wide temperature range and after different thermal histories. The strong dependence on cooling history and measurement time for the microscopic configurations of this PS film showed the non-ergodicity and relaxation behavior. Moreover, the dynamics of glassy behavior, driven and strongly influenced by strain, stop evolving below the freezing temperature. 

The sample is a 40 nm thick PCSMO \cite{c20,c22,c21,c23} film grown on a (110) (LaAlO$_3$)$_{0.3}$(SrAl$_{0.5}$Ta$_{0.5}$O$_3$)$_{0.7}$ (LSAT) substrate by pulse laser deposition. In the X-ray reciprocal maps, the (222) and (310) peaks of the film align with those of the LSAT substrate, indicating a coherent growth (FIG. \ref{fig1} (a)). Atomic force microscopy image of the film shows a flat surface with atomic steps (inset of FIG. \ref{fig1}(a)) and sporadic Mn$_3$O$_4$ precipitates. At zero magnetic ($B$) field, the resistivity measurement shows an insulating behavior at all temperatures (dotted-red curve in FIG. \ref{fig1}(b)), with a transition from PM-I to CO-I at $T_{co} =$  200 K (see the \hyperref[supp]{Supplemental Materials}  \cite{c24} for details). The FM-M state, on the other hand, can be easily recovered by application of a small $B \approx$ 1 T (FIG. \ref{fig1}(b), solid-black). Large hysteresis between the zero field cooling (ZFC) and field warm (FW) curve, noticeable in both resistivity (FIG. \ref{fig1}(b)) and the magnetization curves (\hyperref[supp]{Supplemental Materials}   \cite{c24}), suggests the existence of a PS state. Note that by growing the film on another substrate, CaTiO$_3$/SrTiO$_3$ (CTO/STO) with a lattice constant of 3.826 $\AA$ closer to the bulk PCSMO film than LSAT (3.869 $\AA$) \cite{c25}, the film becomes metallic at low temperature (FIG. \ref{fig1}(b), dash-blue). The transport behavior of PCSMO/LSAT is similar to bulk LPCMO \cite{c6,c7,c8,c17}, where glassy PS state is found. Unlike bulk manganites with accommodation strains \cite{c17}, the PS in epitaxial films is strongly influenced by the strain from the substrate.

In order to identify the glassy PS state, we constructed the $B-T$ phase diagram of the PCSMO film on LSAT using isothermal magnetoresistive curves obtained after ZFC to various temperatures. Examples of such curves at 12, 50 and 130 K are given in FIG. \ref{fig1}(c). For $ T <$ 50 K, the magnetoresistance curves with increasing $B$-fields show a sharp transition to the metallic phase (percolation transition), while the opposite is not seen when sweeping back to zero field (dotted-red and dash-blue curves in FIG. \ref{fig1}(c)). At temperatures higher than 50 K, the transition is smooth and the resistance returns back to the original value with a hysteresis after sweeping back of the fields (solid-green curve).  The $B$-field at the percolation point during the forward sweeping ($B_{per}$) gradually reduces from 12 K to 50 K and increases again at higher temperature. At $T >$ 100 K, the backward sweeping curve shows a return to the high resistance state at $B?_{per}$ . The $B_{per}$ and $B?_{per}$ data points at various temperatures enclose the glass-like region (orange shaded region) on the B-T phase diagram in FIG. \ref{fig1}(d). Outside this region, either the FM-M or CO-I states dominate the system; while inside, a glassy mixture of the FM-M and supercooled CO-I states is expected. In addition, the minimum Bper at 50 K indicates the freezing point of the PS glass \cite{c8}, i.e., the glass transition temperature, $T_g$.

\begin{figure*}

	\centering
		\includegraphics[width=7in]{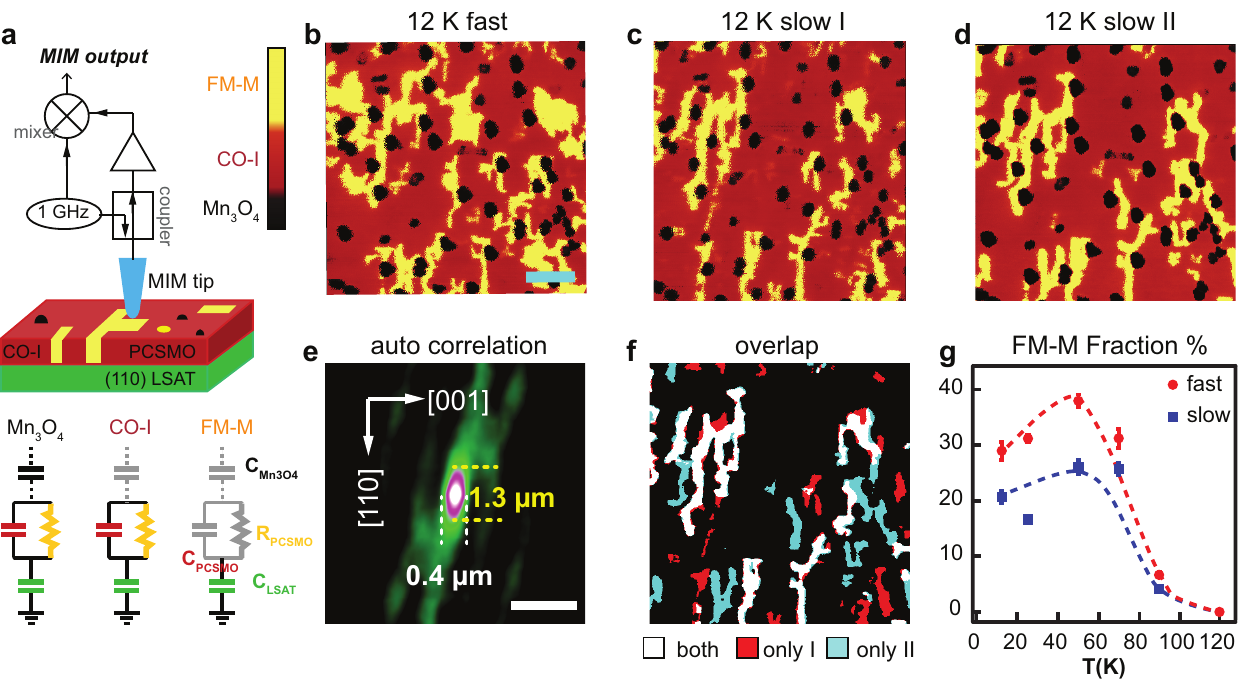}
		\caption{(top) Schematic of the MIM setup and (bottom) the equivalent circuit of the tip-PCSMO film interaction. The MIM responses on the Mn3O4, CO-I and FM-M regions are colored in black, red, and yellow, respectively. (b) ? (d) From left to right, MIM images at 12 K after FCD, and after two SCDs. (e) Autocorrelation analysis of the SCD-I image, illustrating the preferred alignment of FM-M domains along the [1-10] axis of the substrate. (f) Overlay of the domains from the two SCDs. All scale bars are 2 $\mu$m. (g) FM-M fractions of the two processes as a function of temperature.  }
			\label{fig2}
\end{figure*}

Microscopic PS of manganites can be readily observed by MIM \cite{c19}, which functions by sending a 1 GHz microwave signal into a coaxial cantilever probe \cite{c27} and measuring the nanoscale sample impedance. FIG. \ref{fig2}(a) shows a schematic of the cryogenic MIM measurement \cite{c28} (top) and the equivalent lumped-element circuit (bottom). The real and imaginary parts of the MIM outputs contain local conductivity information of the sample, with a spatial resolution down to 50 nm. Mn$_3$O$_4$ precipitates\cite{c29}, which are always insulating within our temperature and field ranges, are present on the sample surface and conveniently serve as landmarks. For this study, the MIM response in the imaginary channel, which is proportional to the tip-sample capacitance \cite{c30}, is sufficient to identify the Mn$_3$O$_4$, CO-I and FM-M regions. We use false color map to label each region as black, red, and yellow, respectively \cite{c31}.

We investigated the low temperature states after cooling at different rates. A fast cool down (FCD) at 8 K/min and a slow cool down (SCD) at 0.3 K/min, were employed to reach specific temperatures at which the images were taken. FIG. \ref{fig2}b and 2c show representative images at 12 K for FCD and SCD, respectively. The substantial FM-M fraction in the nominally insulating film is striking. Indeed, careful image analysis (FIG. \ref{fig2}(g)) shows that the FM-M states can occupy up to 40$\%$ of the sample when cooling at zero field. This result indicates a strong competition between the two states even at zero B-field. Interestingly, the presence of these FM-M clusters is not measurable in either the resistivity (FIG. \ref{fig1}(b)) or the magnetization (\hyperref[supp]{Supplemental Materials}   \cite{c24}). The former is expected for an FM-M faction below the percolation threshold \cite{26}, and the latter is due to the random orientation of macro-spins from the FM-M clusters \cite{32}. 

By considering the crystal structure of each phase (FIG. \ref{fig1}(d)), we can explain how a slow cooling rate leads to fewer FM-M domains. At 200 K, the transition from PM-I to CO-I is accompanied by a structural phase transition \cite{c21,c33}. This process is likely to be highly viscous, and some PM-I regions may directly enter the FM-M rather than the CO-I phase; thus more FM-M domains are found at low temperature under a fast cooling process. Furthermore, in FIG. \ref{fig2}e, the autocorrelation analysis of the SCD images shows the tendency of FM-M domains to elongate along the [1$\bar{1}$0] axis of the substrate, suggesting stronge influences from the local domain structure \cite{c10,c19}. In addition, compared to the metallic behavior in the unstrained film on the CTO/STO substrate, we believe that the PS glassy state in the strained film on a LSAT substrate is mainly induced by substrate elastic strain.

The SCD state has noticeably fewer FM-M domains than that after FCD. Interestingly, while the percentage of area occupied by the FM-M domains are about the same for the two identical SCDs, the MIM images show different FM-M domain configurations. In FIG. \ref{fig2}(f), two SCD images are overlaid for comparison, with about 50$\%$ of the FM-M domains appearing at different locations. The overlapped domains are likely pinned by intrinsic defects \cite{26}, while the random appearance of the rest is a clear evidence of the non-ergodicity.

\begin{figure*}[t!]
	\centering
	\includegraphics[width=7in]{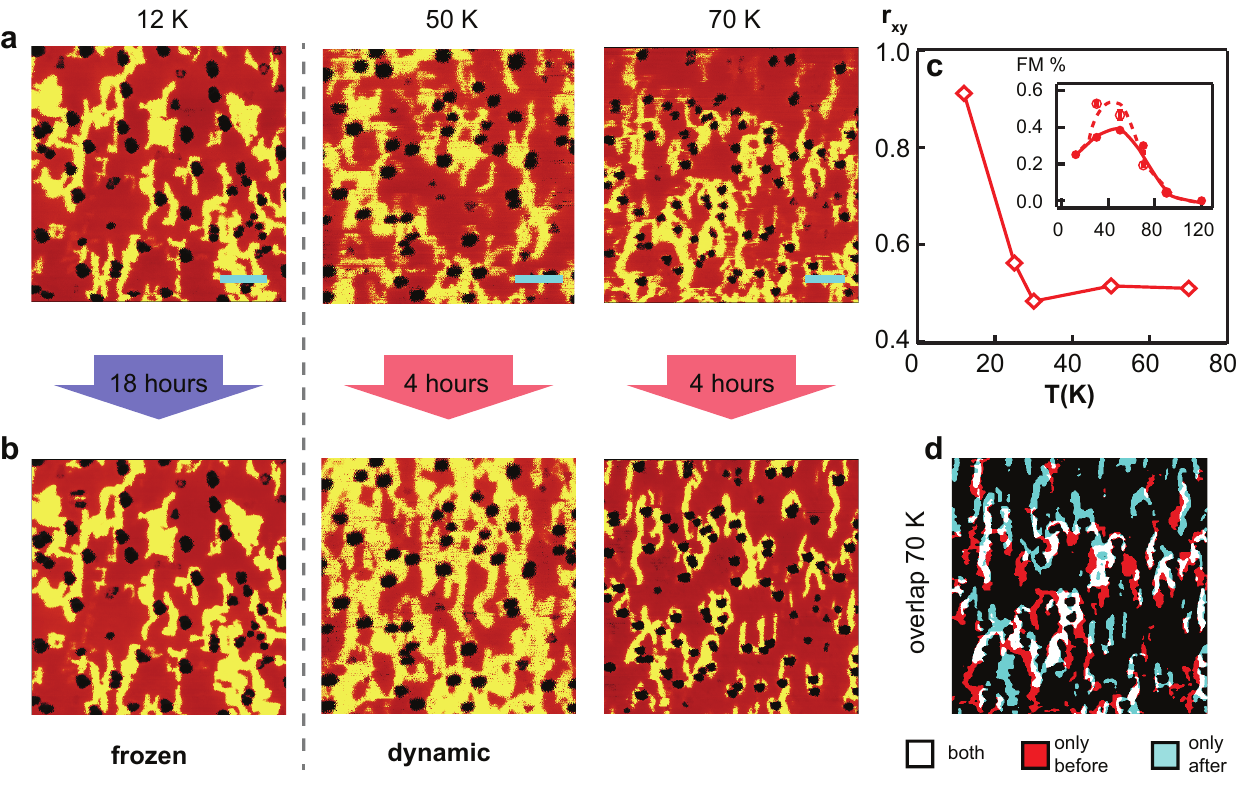}
	\caption{(a) and (b) MIM images after FCD and waiting for several hours at 12, 50 and 70 K, respectively. (c) Cross-correlation coefficient $r_{xy}$  as a function of temperature, showing a sharp jump of $r_{xy}$ at $T_g ~$ 20 K. The inset summarizes the FM-M fractions before (solid) and after (empty circle) a holding time of 4 hours. (d) Overlap of the two images at 70 K, showing the emergence of a number of new FM domains, and the disappearance of previously existing domains.  All scale bars are 2  $\mu$m. }
	\label{fig3}
\end{figure*}

The observed PS states are far from being static, as shown in the relaxation of the PS state after zero-field cooling (ZFC) \cite{c4,c5,c12}. Before each cooling, the sample was warmed up to 250 K, above the charge-order temperature $T_{co}$, to remove the impact of any prior history. The cooling rate was kept at 8 K/min. FIG. \ref{fig3}(a) shows images after ZFC to various temperatures above and below $T_g$. Repeated scans on the same area show drastic changes in the PS at $T >$ 20 K (FIG. \ref{fig3}(b)). After a waiting time of 4 hours at 50 K, the FM-M domains grew from 40$\%$ to 50$\%$. More importantly, many FM-M domains appeared at different locations even though the change in FM-M percentage was small. In other words, a simple measure of the areal fraction (FIG. \ref{fig3}(c) inset) cannot capture the dynamics here. Such dynamical behavior can be illustrated by overlapping the FM-M domains before and after the holding period at 70 K, as shown in FIG. \ref{fig3}(d). On the other hand, the images taken at 12 K show no change after 18 hours. The microscopic rearrangement can be numerically studied by calculating the cross-correlation coefficient $r_{xy}$ between the images before and after the 4-hour interval. As shown in the inset of FIG. \ref{fig3}(c), $r_{xy}$ is small ($\approx$ 0.5) at high temperatures but rapidly rises towards 1 below 20 K, a vivid demonstration of the freezing of the PS glass at this temperature \cite{c8}. 

\begin{figure*}[t!]
	\centering
		\includegraphics[width=7in]{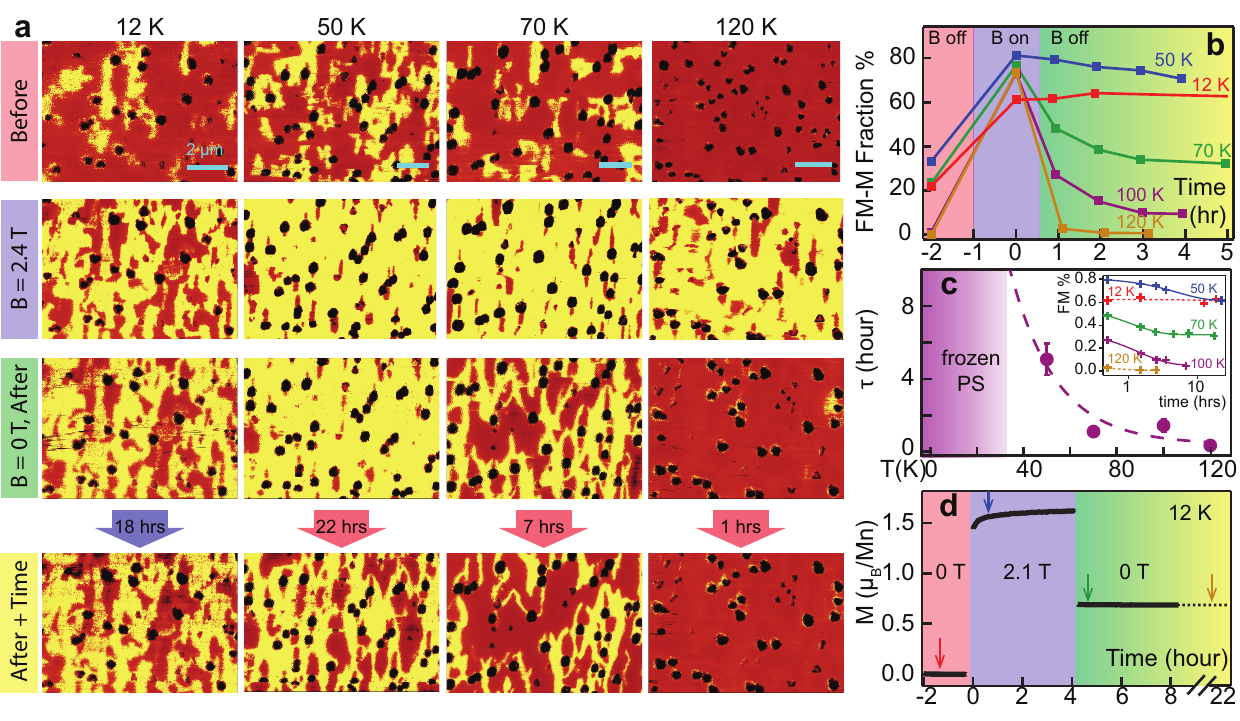}
		\caption{ 
					(a) MIM images taken at$ B =$ 0 T (before), $B =$ 2.4 T, $B = 0$ T (after) and after several hours at the same condition (After + Time) at $T =$ 12, 50, 70, and 120 K, respectively.All scale bars are 2 $\mu$m  (b) FM-M vs time  showing the fraction before, at $B = $2.4 T, turning of the $B$-field at different T. (c) Relaxation time as a function of temperature calculated from an exponential fit of FM-M vs time (d) Magnetization as a function of time. The $B$ field is applied at time zero and removed after 4 hours. The arrows indicate where the MIM images were taken.. }
			\label{fig4}
\end{figure*}

We further investigated the dynamic switching of the PS glass state from the FM-M to CO-I by monitoring the relaxation after removing an external $B$-field \cite{c5} (FIG. \ref{fig4}(a)). The sample was again prepared by ZFC to the desired $T$ before turning on the $B$-field. A field of 2.4 T induced a significant portion of FM-M phases at all temperatures. Immediately after its removal, however, very different behaviors occurred at different temperatures. For $T < T_g$, the PS was frozen with virtually no change in its configuration 18 hours after switching off the field. In contrast, dynamic behavior was observed at higher temperatures. At 50 K, no obvious change was observed right after turning off $B$, while the insulating regions expanded after one day. The relaxation was much faster at 70 K, where large changes happened right after field removal and continued for several hours. For high enough $T =$ 120 K, the relaxation back to the zero-field state was faster than our imaging time, so little variation can be seen after another hour of waiting. FIG. \ref{fig4}(b) summarizes the measured FM-M fraction throughout this process. Using the FM-M fraction as a function of time (FIG. \ref{fig4}(c) inset) and assuming a logarithmic time dependence [4], we can extract the relaxation time as a function of temperature (\hyperref[supp]{Supplemental Materials}   \cite{c24}). As seen in FIG. \ref{fig3}(c), the relaxation time, on the order of several hours, diverges when approaching $T_g$. On the other hand, the magnetization dropped by almost half (FIG. \ref{fig4}(d)) immediately upon field removal even at 12K. In other words, a sizeable amount of spins in the FM-M clusters are still able to randomize even though the cluster itself is frozen. 

In summary, microwave microscopy study of PCSMO film on LSAT reveals a PS texture with non-ergodicity and relaxation behavior, which are the key hallmarks of a glassy state. The glassiness is driven by phase competition and strongly influenced by the tensile strain from the substrate, resulting in preferential alignment of the FM-M domains. The microscopic configurations of PS states highly depend on the cooling history and differ even after the same cooling process. The PS exhibits dynamic behavior such as growing and shrinking of the FM-M domains, but the relaxation virtually stops below the freezing temperature.  The freezing behavior is seen in the rapid growth in cross correlation coefficient from the time dependence images and the relaxation time constant from the ferromagnetic domain areal fraction. By demonstrating a route to visualize and quantify quantum glassy states, this work should facilitate further investigation of electronic phase separation systems.

\section*{Acknowledgements }

We thank Daniel S. Fisher for stimulating discussions. The measurement work done at Stanford University is supported by NSF grants DMR-0906027, the probe development is supported by the Center of Probing the Nanoscale NSF PHY-0425897 and the Gordon and Betty Moore Foundation through Grant GBMF3133 to ZXS. The work done in RIKEN was supported by JSPS FIRST program. 

\bibliographystyle{apsrev} 
\bibliography{referencefile}

\begin{thebibliography}{30}
\expandafter\ifx\csname natexlab\endcsname\relax\def\natexlab#1{#1}\fi
\expandafter\ifx\csname bibnamefont\endcsname\relax
  \def\bibnamefont#1{#1}\fi
\expandafter\ifx\csname bibfnamefont\endcsname\relax
  \def\bibfnamefont#1{#1}\fi
\expandafter\ifx\csname url\endcsname\relax
  \def\url#1{\texttt{#1}}\fi
\expandafter\ifx\csname urlprefix\endcsname\relax\def\urlprefix{URL }\fi
\providecommand*{\bibinfo}[2]{#2}
\providecommand*{\eprint}[1]{#1}
\providecommand*{\url}[1]{#1}
\begingroup\makeatletter
 \@temptokena{%
  \expandafter\ifx\csname citenamefont\endcsname\relax
   \DeclareRobustCommand\citenamefont{\@firstofone}%
   \global\let\citenamefont\citenamefont
   \global\expandafter\let\csname citenamefont \expandafter\endcsname\csname
  citenamefont \endcsname
  \fi
 }\if@filesw\immediate\write\@auxout{\the\@temptokena}\fi
\expandafter\endgroup\the\@temptokena

\bibitem[{\citenamefont{Debenedetti and Stillinger}(2001)}]{c1}
\bibinfo{author}{\bibfnamefont{P.~G.} \bibnamefont{Debenedetti}}
  \bibnamefont{and} \bibinfo{author}{\bibfnamefont{F.~H.}
  \bibnamefont{Stillinger}}, \bibinfo{journal}{Nature}
  \textbf{\bibinfo{volume}{410}}(\bibinfo{number}{6825}), \bibinfo{pages}{259}
  (\bibinfo{year}{2001}), ISSN \bibinfo{issn}{0028-0836}.

\bibitem[{\citenamefont{Dagotto} \emph{et~al.}(2001)\citenamefont{Dagotto,
  Hotta, and Moreo}}]{c2}
\bibinfo{author}{\bibfnamefont{E.}~\bibnamefont{Dagotto}},
  \bibinfo{author}{\bibfnamefont{T.}~\bibnamefont{Hotta}}, \bibnamefont{and}
  \bibinfo{author}{\bibfnamefont{A.}~\bibnamefont{Moreo}},
  \bibinfo{journal}{Physics Reports}
  \textbf{\bibinfo{volume}{344}}(\bibinfo{number}{1-3}), \bibinfo{pages}{1}
  (\bibinfo{year}{2001}), ISSN \bibinfo{issn}{0370-1573}.

\bibitem[{\citenamefont{Dagotto} \emph{et~al.}(2003)\citenamefont{Dagotto,
  Burgy, and Moreo}}]{c3}
\bibinfo{author}{\bibfnamefont{E.}~\bibnamefont{Dagotto}},
  \bibinfo{author}{\bibfnamefont{J.}~\bibnamefont{Burgy}}, \bibnamefont{and}
  \bibinfo{author}{\bibfnamefont{A.}~\bibnamefont{Moreo}},
  \bibinfo{journal}{Solid State Communications}
  \textbf{\bibinfo{volume}{126}}(\bibinfo{number}{1-2}), \bibinfo{pages}{9}
  (\bibinfo{year}{2003}), ISSN \bibinfo{issn}{00381098}, \eprint{0209689}.

\bibitem[{\citenamefont{Sirena} \emph{et~al.}(2001)\citenamefont{Sirena,
  Steren, and Guimpel}}]{c4}
\bibinfo{author}{\bibfnamefont{M.}~\bibnamefont{Sirena}},
  \bibinfo{author}{\bibfnamefont{L.~B.} \bibnamefont{Steren}},
  \bibnamefont{and} \bibinfo{author}{\bibfnamefont{J.}~\bibnamefont{Guimpel}},
  \bibinfo{journal}{Physical Review B}
  \textbf{\bibinfo{volume}{64}}(\bibinfo{number}{10}), \bibinfo{pages}{104409}
  (\bibinfo{year}{2001}), ISSN \bibinfo{issn}{0163-1829}.

\bibitem[{\citenamefont{L{\'{o}}pez}
  \emph{et~al.}(2001)\citenamefont{L{\'{o}}pez, Lisboa-Filho, Passos, Ortiz,
  Araujo-Moreira, de~Lima, Schaniel, and Ghosh}}]{c5}
\bibinfo{author}{\bibfnamefont{J.}~\bibnamefont{L{\'{o}}pez}},
  \bibinfo{author}{\bibfnamefont{P.}~\bibnamefont{Lisboa-Filho}},
  \bibinfo{author}{\bibfnamefont{W.}~\bibnamefont{Passos}},
  \bibinfo{author}{\bibfnamefont{W.}~\bibnamefont{Ortiz}},
  \bibinfo{author}{\bibfnamefont{F.}~\bibnamefont{Araujo-Moreira}},
  \bibinfo{author}{\bibfnamefont{O.}~\bibnamefont{de~Lima}},
  \bibinfo{author}{\bibfnamefont{D.}~\bibnamefont{Schaniel}}, \bibnamefont{and}
  \bibinfo{author}{\bibfnamefont{K.}~\bibnamefont{Ghosh}},
  \bibinfo{journal}{Physical Review B}
  \textbf{\bibinfo{volume}{63}}(\bibinfo{number}{22}), \bibinfo{pages}{224422}
  (\bibinfo{year}{2001}), ISSN \bibinfo{issn}{0163-1829}.

\bibitem[{\citenamefont{Ghivelder and Parisi}(2005)}]{c6}
\bibinfo{author}{\bibfnamefont{L.}~\bibnamefont{Ghivelder}} \bibnamefont{and}
  \bibinfo{author}{\bibfnamefont{F.}~\bibnamefont{Parisi}},
  \bibinfo{journal}{Physical Review B}
  \textbf{\bibinfo{volume}{71}}(\bibinfo{number}{18}), \bibinfo{pages}{184425}
  (\bibinfo{year}{2005}), ISSN \bibinfo{issn}{1098-0121}.

\bibitem[{\citenamefont{Sharma} \emph{et~al.}(2005)\citenamefont{Sharma, Kim,
  Koo, Guha, and Cheong}}]{c7}
\bibinfo{author}{\bibfnamefont{P.}~\bibnamefont{Sharma}},
  \bibinfo{author}{\bibfnamefont{S.}~\bibnamefont{Kim}},
  \bibinfo{author}{\bibfnamefont{T.}~\bibnamefont{Koo}},
  \bibinfo{author}{\bibfnamefont{S.}~\bibnamefont{Guha}}, \bibnamefont{and}
  \bibinfo{author}{\bibfnamefont{S.-W.} \bibnamefont{Cheong}},
  \bibinfo{journal}{Physical Review B}
  \textbf{\bibinfo{volume}{71}}(\bibinfo{number}{22}), \bibinfo{pages}{224416}
  (\bibinfo{year}{2005}), ISSN \bibinfo{issn}{1098-0121}.

\bibitem[{\citenamefont{Sharma} \emph{et~al.}(2008)\citenamefont{Sharma,
  El-Khatib, Mihut, Betts, Migliori, Kim, Guha, and Cheong}}]{c8}
\bibinfo{author}{\bibfnamefont{P.~a.} \bibnamefont{Sharma}},
  \bibinfo{author}{\bibfnamefont{S.}~\bibnamefont{El-Khatib}},
  \bibinfo{author}{\bibfnamefont{I.}~\bibnamefont{Mihut}},
  \bibinfo{author}{\bibfnamefont{J.~B.} \bibnamefont{Betts}},
  \bibinfo{author}{\bibfnamefont{a.}~\bibnamefont{Migliori}},
  \bibinfo{author}{\bibfnamefont{S.~B.} \bibnamefont{Kim}},
  \bibinfo{author}{\bibfnamefont{S.}~\bibnamefont{Guha}}, \bibnamefont{and}
  \bibinfo{author}{\bibfnamefont{S.-W.~W.} \bibnamefont{Cheong}},
  \bibinfo{journal}{Physical Review B}
  \textbf{\bibinfo{volume}{78}}(\bibinfo{number}{13}), \bibinfo{pages}{134205}
  (\bibinfo{year}{2008}), ISSN \bibinfo{issn}{1098-0121}.

\bibitem[{\citenamefont{Tomioka and Tokura}(2004)}]{c9}
\bibinfo{author}{\bibfnamefont{Y.}~\bibnamefont{Tomioka}} \bibnamefont{and}
  \bibinfo{author}{\bibfnamefont{Y.}~\bibnamefont{Tokura}},
  \bibinfo{journal}{Physical Review B}
  \textbf{\bibinfo{volume}{70}}(\bibinfo{number}{1}), \bibinfo{pages}{1}
  (\bibinfo{year}{2004}), ISSN \bibinfo{issn}{1098-0121}.

\bibitem[{\citenamefont{Ahn} \emph{et~al.}(2004)\citenamefont{Ahn, Lookman, and
  Bishop}}]{c10}
\bibinfo{author}{\bibfnamefont{K.~H.} \bibnamefont{Ahn}},
  \bibinfo{author}{\bibfnamefont{T.}~\bibnamefont{Lookman}}, \bibnamefont{and}
  \bibinfo{author}{\bibfnamefont{A.~R.} \bibnamefont{Bishop}}
  \textbf{\bibinfo{volume}{804}}(\bibinfo{number}{December 2003}),
  \bibinfo{pages}{401} (\bibinfo{year}{2004}).

\bibitem[{\citenamefont{Sacanell} \emph{et~al.}(2006)\citenamefont{Sacanell,
  Parisi, Campoy, and Ghivelder}}]{c11}
\bibinfo{author}{\bibfnamefont{J.}~\bibnamefont{Sacanell}},
  \bibinfo{author}{\bibfnamefont{F.}~\bibnamefont{Parisi}},
  \bibinfo{author}{\bibfnamefont{J.}~\bibnamefont{Campoy}}, \bibnamefont{and}
  \bibinfo{author}{\bibfnamefont{L.}~\bibnamefont{Ghivelder}},
  \bibinfo{journal}{Physical Review B}
  \textbf{\bibinfo{volume}{73}}(\bibinfo{number}{1}), \bibinfo{pages}{014403}
  (\bibinfo{year}{2006}), ISSN \bibinfo{issn}{1098-0121}.

\bibitem[{\citenamefont{Deac} \emph{et~al.}(2002)\citenamefont{Deac, Diaz, Kim,
  Cheong, and Schiffer}}]{c12}
\bibinfo{author}{\bibfnamefont{I.}~\bibnamefont{Deac}},
  \bibinfo{author}{\bibfnamefont{S.}~\bibnamefont{Diaz}},
  \bibinfo{author}{\bibfnamefont{B.}~\bibnamefont{Kim}},
  \bibinfo{author}{\bibfnamefont{S.-W.} \bibnamefont{Cheong}},
  \bibnamefont{and} \bibinfo{author}{\bibfnamefont{P.}~\bibnamefont{Schiffer}},
  \bibinfo{journal}{Physical Review B}
  \textbf{\bibinfo{volume}{65}}(\bibinfo{number}{17}), \bibinfo{pages}{174426}
  (\bibinfo{year}{2002}), ISSN \bibinfo{issn}{0163-1829}.

\bibitem[{\citenamefont{Ward} \emph{et~al.}(2011)\citenamefont{Ward, Gai, Guo,
  Yin, and Shen}}]{c13}
\bibinfo{author}{\bibfnamefont{T.~Z.} \bibnamefont{Ward}},
  \bibinfo{author}{\bibfnamefont{Z.}~\bibnamefont{Gai}},
  \bibinfo{author}{\bibfnamefont{H.~W.} \bibnamefont{Guo}},
  \bibinfo{author}{\bibfnamefont{L.~F.} \bibnamefont{Yin}}, \bibnamefont{and}
  \bibinfo{author}{\bibfnamefont{J.}~\bibnamefont{Shen}},
  \bibinfo{journal}{Physical Review B - Condensed Matter and Materials Physics}
  \textbf{\bibinfo{volume}{83}}(\bibinfo{number}{12}), \bibinfo{pages}{1}
  (\bibinfo{year}{2011}), ISSN \bibinfo{issn}{10980121}.

\bibitem[{\citenamefont{F{\"{a}}th}
  \emph{et~al.}(1999)\citenamefont{F{\"{a}}th, Freisem, Menovsky, Tomioka,
  Aarts, and Mydosh}}]{c14}
\bibinfo{author}{\bibfnamefont{M.}~\bibnamefont{F{\"{a}}th}},
  \bibinfo{author}{\bibfnamefont{S.}~\bibnamefont{Freisem}},
  \bibinfo{author}{\bibfnamefont{A.~A.} \bibnamefont{Menovsky}},
  \bibinfo{author}{\bibfnamefont{Y.}~\bibnamefont{Tomioka}},
  \bibinfo{author}{\bibfnamefont{J.}~\bibnamefont{Aarts}}, \bibnamefont{and}
  \bibinfo{author}{\bibfnamefont{J.~A.} \bibnamefont{Mydosh}},
  \bibinfo{journal}{Science}
  \textbf{\bibinfo{volume}{285}}(\bibinfo{number}{5433}), \bibinfo{pages}{1540}
  (\bibinfo{year}{1999}).

\bibitem[{\citenamefont{He} \emph{et~al.}(2010)\citenamefont{He, Volkov, Asaka,
  Chaudhuri, Budhani, and Zhu}}]{c15}
\bibinfo{author}{\bibfnamefont{J.~Q.} \bibnamefont{He}},
  \bibinfo{author}{\bibfnamefont{V.~V.} \bibnamefont{Volkov}},
  \bibinfo{author}{\bibfnamefont{T.}~\bibnamefont{Asaka}},
  \bibinfo{author}{\bibfnamefont{S.}~\bibnamefont{Chaudhuri}},
  \bibinfo{author}{\bibfnamefont{R.~C.} \bibnamefont{Budhani}},
  \bibnamefont{and} \bibinfo{author}{\bibfnamefont{Y.}~\bibnamefont{Zhu}},
  \bibinfo{journal}{Physical Review B}
  \textbf{\bibinfo{volume}{82}}(\bibinfo{number}{22}), \bibinfo{pages}{224404}
  (\bibinfo{year}{2010}).

\bibitem[{\citenamefont{Burkhardt} \emph{et~al.}(2012)\citenamefont{Burkhardt,
  Hossain, Sarkar, Chuang, {Cruz Gonzalez}, Doran, Scholl, Young, Tahir, Choi,
  Cheong, D{\"{u}}rr} \emph{et~al.}}]{c16}
\bibinfo{author}{\bibfnamefont{M.}~\bibnamefont{Burkhardt}},
  \bibinfo{author}{\bibfnamefont{M.}~\bibnamefont{Hossain}},
  \bibinfo{author}{\bibfnamefont{S.}~\bibnamefont{Sarkar}},
  \bibinfo{author}{\bibfnamefont{Y.-D.} \bibnamefont{Chuang}},
  \bibinfo{author}{\bibfnamefont{A.}~\bibnamefont{{Cruz Gonzalez}}},
  \bibinfo{author}{\bibfnamefont{A.}~\bibnamefont{Doran}},
  \bibinfo{author}{\bibfnamefont{A.}~\bibnamefont{Scholl}},
  \bibinfo{author}{\bibfnamefont{A.}~\bibnamefont{Young}},
  \bibinfo{author}{\bibfnamefont{N.}~\bibnamefont{Tahir}},
  \bibinfo{author}{\bibfnamefont{Y.}~\bibnamefont{Choi}},
  \bibinfo{author}{\bibfnamefont{S.-W.} \bibnamefont{Cheong}},
  \bibinfo{author}{\bibfnamefont{H.}~\bibnamefont{D{\"{u}}rr}}, \emph{et~al.},
  \bibinfo{journal}{Physical Review Letters}
  \textbf{\bibinfo{volume}{108}}(\bibinfo{number}{23}), \bibinfo{pages}{237202}
  (\bibinfo{year}{2012}), ISSN \bibinfo{issn}{0031-9007}.

\bibitem[{\citenamefont{Wu} \emph{et~al.}(2006)\citenamefont{Wu, Israel, Hur,
  Park, Cheong, and de~Lozanne}}]{c17}
\bibinfo{author}{\bibfnamefont{W.}~\bibnamefont{Wu}},
  \bibinfo{author}{\bibfnamefont{C.}~\bibnamefont{Israel}},
  \bibinfo{author}{\bibfnamefont{N.}~\bibnamefont{Hur}},
  \bibinfo{author}{\bibfnamefont{S.}~\bibnamefont{Park}},
  \bibinfo{author}{\bibfnamefont{S.-W.} \bibnamefont{Cheong}},
  \bibnamefont{and}
  \bibinfo{author}{\bibfnamefont{A.}~\bibnamefont{de~Lozanne}},
  \bibinfo{journal}{Nat Mater}
  \textbf{\bibinfo{volume}{5}}(\bibinfo{number}{11}), \bibinfo{pages}{881}
  (\bibinfo{year}{2006}), ISSN \bibinfo{issn}{1476-1122}.

\bibitem[{\citenamefont{Rawat} \emph{et~al.}(2013)\citenamefont{Rawat,
  Kushwaha, Mishra, and Sathe}}]{c18}
\bibinfo{author}{\bibfnamefont{R.}~\bibnamefont{Rawat}},
  \bibinfo{author}{\bibfnamefont{P.}~\bibnamefont{Kushwaha}},
  \bibinfo{author}{\bibfnamefont{D.~K.} \bibnamefont{Mishra}},
  \bibnamefont{and} \bibinfo{author}{\bibfnamefont{V.~G.} \bibnamefont{Sathe}},
  \bibinfo{journal}{Physical Review B}
  \textbf{\bibinfo{volume}{87}}(\bibinfo{number}{6}), \bibinfo{pages}{064412}
  (\bibinfo{year}{2013}), ISSN \bibinfo{issn}{1098-0121}.

\bibitem[{\citenamefont{Lai} \emph{et~al.}(2010)\citenamefont{Lai, Nakamura,
  Kundhikanjana, Kawasaki, Tokura, Kelly, and Shen}}]{c19}
\bibinfo{author}{\bibfnamefont{K.}~\bibnamefont{Lai}},
  \bibinfo{author}{\bibfnamefont{M.}~\bibnamefont{Nakamura}},
  \bibinfo{author}{\bibfnamefont{W.}~\bibnamefont{Kundhikanjana}},
  \bibinfo{author}{\bibfnamefont{M.}~\bibnamefont{Kawasaki}},
  \bibinfo{author}{\bibfnamefont{Y.}~\bibnamefont{Tokura}},
  \bibinfo{author}{\bibfnamefont{M.~A.} \bibnamefont{Kelly}}, \bibnamefont{and}
  \bibinfo{author}{\bibfnamefont{Z.-X.} \bibnamefont{Shen}},
  \bibinfo{journal}{Science}
  \textbf{\bibinfo{volume}{329}}(\bibinfo{number}{5988}), \bibinfo{pages}{190}
  (\bibinfo{year}{2010}).

\bibitem[{\citenamefont{Takubo} \emph{et~al.}(2005)\citenamefont{Takubo,
  Ogimoto, Nakamura, Tamaru, Izumi, and Miyano}}]{c20}
\bibinfo{author}{\bibfnamefont{N.}~\bibnamefont{Takubo}},
  \bibinfo{author}{\bibfnamefont{Y.}~\bibnamefont{Ogimoto}},
  \bibinfo{author}{\bibfnamefont{M.}~\bibnamefont{Nakamura}},
  \bibinfo{author}{\bibfnamefont{H.}~\bibnamefont{Tamaru}},
  \bibinfo{author}{\bibfnamefont{M.}~\bibnamefont{Izumi}}, \bibnamefont{and}
  \bibinfo{author}{\bibfnamefont{K.}~\bibnamefont{Miyano}},
  \bibinfo{journal}{Physical Review Letters}
  \textbf{\bibinfo{volume}{95}}(\bibinfo{number}{1}), \bibinfo{pages}{017404}
  (\bibinfo{year}{2005}), ISSN \bibinfo{issn}{0031-9007}.

\bibitem[{\citenamefont{Sheng} \emph{et~al.}(2012)\citenamefont{Sheng,
  Nakamura, Kagawa, Kawasaki, and Tokura}}]{c22}
\bibinfo{author}{\bibfnamefont{Z.}~\bibnamefont{Sheng}},
  \bibinfo{author}{\bibfnamefont{M.}~\bibnamefont{Nakamura}},
  \bibinfo{author}{\bibfnamefont{F.}~\bibnamefont{Kagawa}},
  \bibinfo{author}{\bibfnamefont{M.}~\bibnamefont{Kawasaki}}, \bibnamefont{and}
  \bibinfo{author}{\bibfnamefont{Y.}~\bibnamefont{Tokura}},
  \bibinfo{journal}{Nature communications}
  \textbf{\bibinfo{volume}{3}}(\bibinfo{number}{May}), \bibinfo{pages}{944}
  (\bibinfo{year}{2012}), ISSN \bibinfo{issn}{2041-1723}.

\bibitem[{\citenamefont{Wakabayashi}
  \emph{et~al.}(2009)\citenamefont{Wakabayashi, Sagayama, Arima, Nakamura,
  Ogimoto, Kubo, Miyano, and Sawa}}]{c21}
\bibinfo{author}{\bibfnamefont{Y.}~\bibnamefont{Wakabayashi}},
  \bibinfo{author}{\bibfnamefont{H.}~\bibnamefont{Sagayama}},
  \bibinfo{author}{\bibfnamefont{T.}~\bibnamefont{Arima}},
  \bibinfo{author}{\bibfnamefont{M.}~\bibnamefont{Nakamura}},
  \bibinfo{author}{\bibfnamefont{Y.}~\bibnamefont{Ogimoto}},
  \bibinfo{author}{\bibfnamefont{Y.}~\bibnamefont{Kubo}},
  \bibinfo{author}{\bibfnamefont{K.}~\bibnamefont{Miyano}}, \bibnamefont{and}
  \bibinfo{author}{\bibfnamefont{H.}~\bibnamefont{Sawa}},
  \bibinfo{journal}{Physical Review B}
  \textbf{\bibinfo{volume}{79}}(\bibinfo{number}{22}), \bibinfo{pages}{2}
  (\bibinfo{year}{2009}), ISSN \bibinfo{issn}{1098-0121}.

\bibitem[{\citenamefont{Okuyama} \emph{et~al.}(2009)\citenamefont{Okuyama,
  Nakamura, Wakabayashi, Itoh, Kumai, Yamada, Taguchi, Arima, Kawasaki, and
  Tokura}}]{c23}
\bibinfo{author}{\bibfnamefont{D.}~\bibnamefont{Okuyama}},
  \bibinfo{author}{\bibfnamefont{M.}~\bibnamefont{Nakamura}},
  \bibinfo{author}{\bibfnamefont{Y.}~\bibnamefont{Wakabayashi}},
  \bibinfo{author}{\bibfnamefont{H.}~\bibnamefont{Itoh}},
  \bibinfo{author}{\bibfnamefont{R.}~\bibnamefont{Kumai}},
  \bibinfo{author}{\bibfnamefont{H.}~\bibnamefont{Yamada}},
  \bibinfo{author}{\bibfnamefont{Y.}~\bibnamefont{Taguchi}},
  \bibinfo{author}{\bibfnamefont{T.}~\bibnamefont{Arima}},
  \bibinfo{author}{\bibfnamefont{M.}~\bibnamefont{Kawasaki}}, \bibnamefont{and}
  \bibinfo{author}{\bibfnamefont{Y.}~\bibnamefont{Tokura}},
  \bibinfo{journal}{Applied Physics Letters}
  \textbf{\bibinfo{volume}{95}}(\bibinfo{number}{15}), \bibinfo{pages}{152502}
  (\bibinfo{year}{2009}), ISSN \bibinfo{issn}{00036951}.

\bibitem[{c24()}]{c24}
\bibinfo{journal}{Supplemental Materials}  (????).

\bibitem[{\citenamefont{Tomioka and Tokura}(2002)}]{c25}
\bibinfo{author}{\bibfnamefont{Y.}~\bibnamefont{Tomioka}} \bibnamefont{and}
  \bibinfo{author}{\bibfnamefont{Y.}~\bibnamefont{Tokura}},
  \bibinfo{journal}{Physical Review B}
  \textbf{\bibinfo{volume}{66}}(\bibinfo{number}{10}), \bibinfo{pages}{104416}
  (\bibinfo{year}{2002}), ISSN \bibinfo{issn}{0163-1829}.

\bibitem[{\citenamefont{Yang} \emph{et~al.}(2012)\citenamefont{Yang, Lai, Tang,
  Kundhikanjana, Kelly, Zhang, Shen, and Li}}]{c27}
\bibinfo{author}{\bibfnamefont{Y.}~\bibnamefont{Yang}},
  \bibinfo{author}{\bibfnamefont{K.}~\bibnamefont{Lai}},
  \bibinfo{author}{\bibfnamefont{Q.}~\bibnamefont{Tang}},
  \bibinfo{author}{\bibfnamefont{W.}~\bibnamefont{Kundhikanjana}},
  \bibinfo{author}{\bibfnamefont{M.~A.} \bibnamefont{Kelly}},
  \bibinfo{author}{\bibfnamefont{K.}~\bibnamefont{Zhang}},
  \bibinfo{author}{\bibfnamefont{Z.-x.~X.} \bibnamefont{Shen}},
  \bibnamefont{and} \bibinfo{author}{\bibfnamefont{X.}~\bibnamefont{Li}},
  \bibinfo{journal}{Journal of Micromechanics and Microengineering}
  \textbf{\bibinfo{volume}{22}}(\bibinfo{number}{11}), \bibinfo{pages}{115040}
  (\bibinfo{year}{2012}).

\bibitem[{\citenamefont{Kundhikanjana}
  \emph{et~al.}(2011)\citenamefont{Kundhikanjana, Lai, Kelly, and Shen}}]{c28}
\bibinfo{author}{\bibfnamefont{W.}~\bibnamefont{Kundhikanjana}},
  \bibinfo{author}{\bibfnamefont{K.}~\bibnamefont{Lai}},
  \bibinfo{author}{\bibfnamefont{M.~A.} \bibnamefont{Kelly}}, \bibnamefont{and}
  \bibinfo{author}{\bibfnamefont{Z.-X.} \bibnamefont{Shen}},
  \bibinfo{journal}{Review of Scientific Instruments}
  \textbf{\bibinfo{volume}{82}}(\bibinfo{number}{3}), \bibinfo{pages}{033705}
  (\bibinfo{year}{2011}), ISSN \bibinfo{issn}{00346748}.

\bibitem[{\citenamefont{Higuchi} \emph{et~al.}(2009)\citenamefont{Higuchi,
  Yajima, Kourkoutis, Hikita, Nakagawa, Muller, and Hwang}}]{c29}
\bibinfo{author}{\bibfnamefont{T.}~\bibnamefont{Higuchi}},
  \bibinfo{author}{\bibfnamefont{T.}~\bibnamefont{Yajima}},
  \bibinfo{author}{\bibfnamefont{L.~F.} \bibnamefont{Kourkoutis}},
  \bibinfo{author}{\bibfnamefont{Y.}~\bibnamefont{Hikita}},
  \bibinfo{author}{\bibfnamefont{N.}~\bibnamefont{Nakagawa}},
  \bibinfo{author}{\bibfnamefont{D.~a.} \bibnamefont{Muller}},
  \bibnamefont{and} \bibinfo{author}{\bibfnamefont{H.~Y.} \bibnamefont{Hwang}},
  \bibinfo{journal}{Applied Physics Letters}
  \textbf{\bibinfo{volume}{95}}(\bibinfo{number}{4}), \bibinfo{pages}{043112}
  (\bibinfo{year}{2009}), ISSN \bibinfo{issn}{00036951}.

\bibitem[{\citenamefont{Lai} \emph{et~al.}(2008)\citenamefont{Lai,
  Kundhikanjana, Kelly, and Shen}}]{c30}
\bibinfo{author}{\bibfnamefont{K.}~\bibnamefont{Lai}},
  \bibinfo{author}{\bibfnamefont{W.}~\bibnamefont{Kundhikanjana}},
  \bibinfo{author}{\bibfnamefont{M.~A.} \bibnamefont{Kelly}}, \bibnamefont{and}
  \bibinfo{author}{\bibfnamefont{Z.~X.} \bibnamefont{Shen}},
  \bibinfo{journal}{The Review of Scientific Instruments}
  \textbf{\bibinfo{volume}{79}}(\bibinfo{number}{6}), \bibinfo{pages}{063703}
  (\bibinfo{year}{2008}), ISSN \bibinfo{issn}{0034-6748}.

\bibitem[{\citenamefont{Wakabayashi} \emph{et~al.}()\citenamefont{Wakabayashi,
  Sawa, Takubo, Nakamura, Ogimoto, and Miyano}}]{c33}
\bibinfo{author}{\bibfnamefont{Y.}~\bibnamefont{Wakabayashi}},
  \bibinfo{author}{\bibfnamefont{H.}~\bibnamefont{Sawa}},
  \bibinfo{author}{\bibfnamefont{N.}~\bibnamefont{Takubo}},
  \bibinfo{author}{\bibfnamefont{M.}~\bibnamefont{Nakamura}},
  \bibinfo{author}{\bibfnamefont{Y.}~\bibnamefont{Ogimoto}}, \bibnamefont{and}
  \bibinfo{author}{\bibfnamefont{K.}~\bibnamefont{Miyano}}  (????).

\end{thebibliography}

\newpage

\section*{Supplemental Materials}
\label{supp}
\renewcommand{\thesubsection}{S\arabic{subsection}}
\renewcommand{\thefigure}{S\arabic{figure}}
\setcounter{figure}{0}

\subsection{PCSMO films on LSAT (110) and CTO/STO (110) Comparision}

\begin{figure}[t]
	\centering
		\includegraphics[width=3.5in]{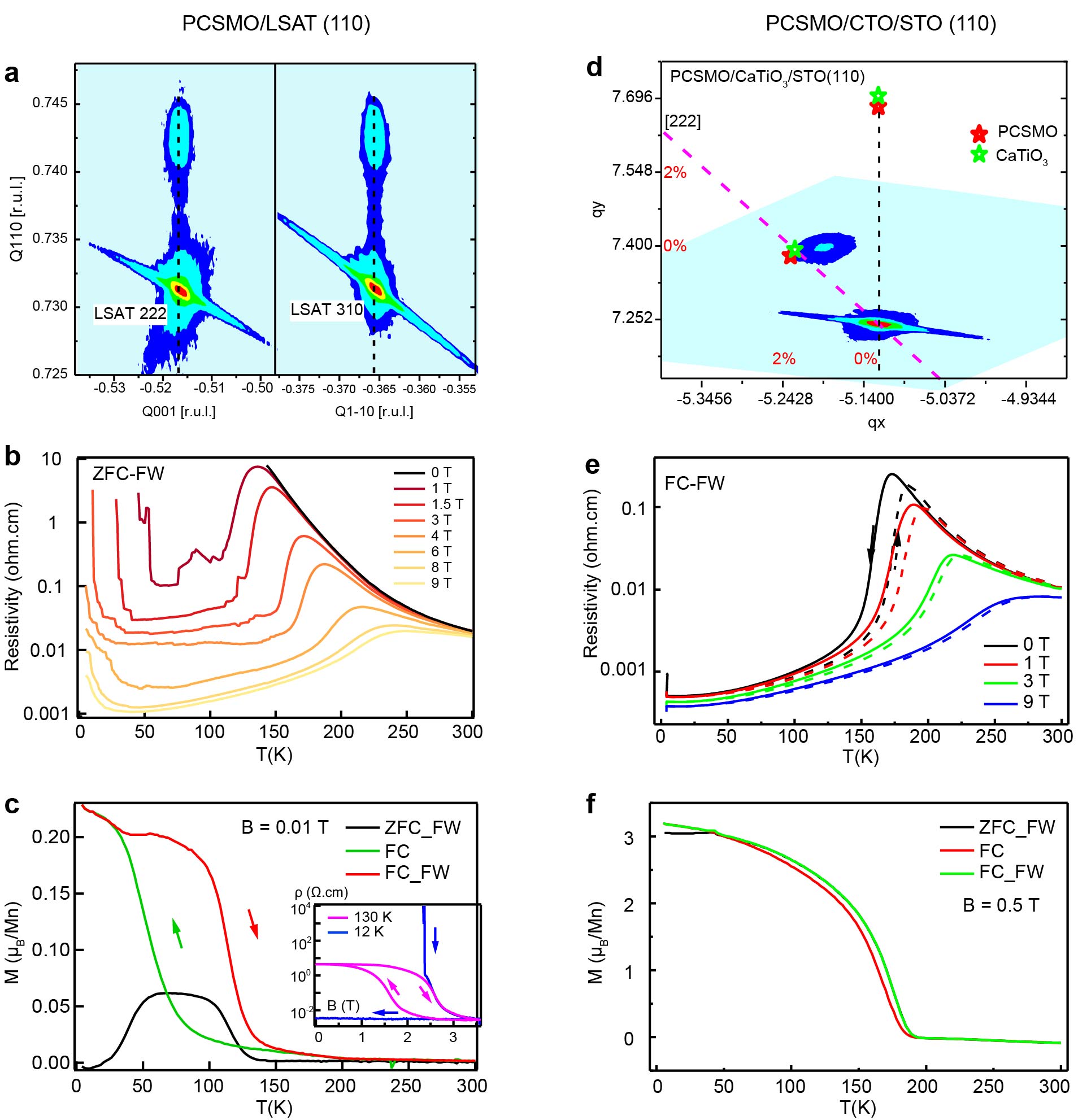}
		\caption{Characterization of PCSMO films on LSAT and CTO/STO (110) substrates }
			\label{sfig1}
\end{figure}

Two high qualities Pr0.55(Ca1-ySry)0.45MnO3 (PCSMO) with y = 0.25 films growth on LSAT and CTO/STO substrates are characterizing by X-ray reciprocal mapping, resistivity and magnetization measurements. FIG. \ref{sfig1}(a) is reciprocal space map of the PCSMO film on LSAT. The (222) and (310) peaks of the film align with the LSAT substrate. The widths of diffraction peaks of film along lateral direction, corresponding to the omega scan of crystal, are comparable to those of substrates suggesting a coherent growth and high quality film. In FIG. \ref{sfig1}(b), the ZFC-FW resistivity curves in different magnetic fields show the reentrance of the insulating state, similar to another bi-critical phase manganite (La,Pr)$_{0.625}$Ca$_{0.375}$MnO$_3$ (For example, PRB \textbf{71}, 224416(2005)). \ref{sfig1}(c), magnetization as a function of temperature for ZFC-FW (black), FC (green) and, FC-FW (red) at 0.01 T. The magnetization was negligible during the ZFC, but could go up to 3.5 $\mu_B$/Mn during the FW if magnetic field of 5 T was used.  PCSMO film on LSAT substrate is quite different from a film on CTO/STO and a bulk sample.

The PCSMO film on CTO/STO substrate behaves rather similar to the bulk sample. FIG.  \ref{sfig1}(d) shows reciprocal space map of the PCSMO film on CTO/STO shows the large deviation from that on the LSAT substrate. Further analysis indicates that the film on CTO/STO is almost relaxed and lattice parameter is quite close to that of bulk case. The resistivity measurements in FIG.  \ref{sfig1}(e) indicate that the film on the CTO/STO substrate is metallic at low temperature. FIG.  \ref{sfig1}(f), Magnetization curve taken during ZFC-FW, FC, FC-FW are the same and no reentrance behavior is observed. The absence of the insulating behavior in the CTO/STO film indicates the influence of the substrate on the properties. The lattice constant of CTO in a cubic convention is 3.826 $\AA$, closer to that of the PCSMO lattice (PRB \textbf{66}, 104416 (2002)), while the lattice constant of LSAT is larger, 3.869 $\AA$. We called the film on CTO/STO a relaxed film and on LSAT a strained film. The results shown above imply that the tensile strain from the LSAT substrate favors the formation of CO-I phase in PCSMO films with FM-M matrix. The coexistence and competition of and the CO-I and FM-M phases in strained film might be responsible for the existence of glassy behavior.

\subsection{Properties of PCSMO film on LSAT}

\begin{figure}
	\centering
		\includegraphics[width=3in]{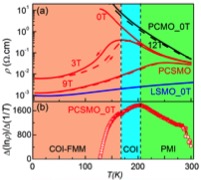}
		\caption{(a) Temperature dependence of resistivity ($\rho$) in various magnetic fields for the PCSMO film on LSAT. (b) $\Delta(ln \rho)/\Delta(1/T)$ is plotted as a function of temperature. }
			\label{sfig2}
\end{figure}

This section focuses on the properties of the PCSMO film on LSAT. FIG. \ref{sfig2}(a) shows resistivity($\rho$) as a function of temperature taken during the FC-FW cycle at B = 0, 3, and 9 T. Cooling the sample under the magnetic field induced the FM-M states at low temperature. Transition from PM-I to CO-I at high temperature is less obvious in the PCSMO film than in the bulk. To identify the transition point, we plot $\Delta(ln \rho)/\Delta(1/T)$ as a function of temperature (APL \textbf{95},152502 (2009)) in FIG. \ref{sfig2}(b). The anomaly at the PM-I to CO-I transition is seen around 200 K.

\begin{figure}
	\centering
	\includegraphics[width=3.5in]{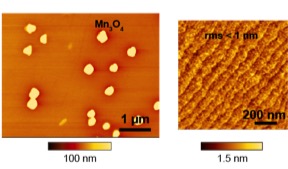}
	\caption{PCSMO films on LSAT substrates showing atomic step terrace between the Mn3O4 particles.}
	\label{sfig3}
\end{figure}

FIG. \ref{sfig3} shows AFM images of the strained PCSMO film. There are Mn$_3$O$_4$ precipitates on the film surface and the regions between the precipitates are flat with clear step and terrace structures. Similar precipitates were also seen in the film on CTO/STO substrate. The presence of Mn$_3$O$_4$ particles implies neither off-stoichiometry films nor different stoichiometry between the films on different substrates. As shown in FIG. \ref{sfig1}(a), the PCSMO films are well clamped onto the LSAT substrate and the narrow diffraction point indicates the high crystalline quality of our films. Moreover, the flat surface with clear step and terrace structures between the precipitates also suggests the high crystal quality of strain films (FIG. \ref{sfig3}). The PCSMO films on both LSAT and CTO/STO were grown with same conditions with similar topographical features. The relaxed PCSMO film on CTO/STO substrate, despite some precipitates, behaves similarly to the bulk indicating correct stoichiometry. 

\subsection{Image analysis}

This section explains the procedure for extracting the areal fraction of the FM-M domains, performing the autocorrelation analysis, overlaying two MIM images, and calculating the cross correlation coefficient. 

\begin{figure}[!]
	\centering
	\includegraphics[width=3.5in]{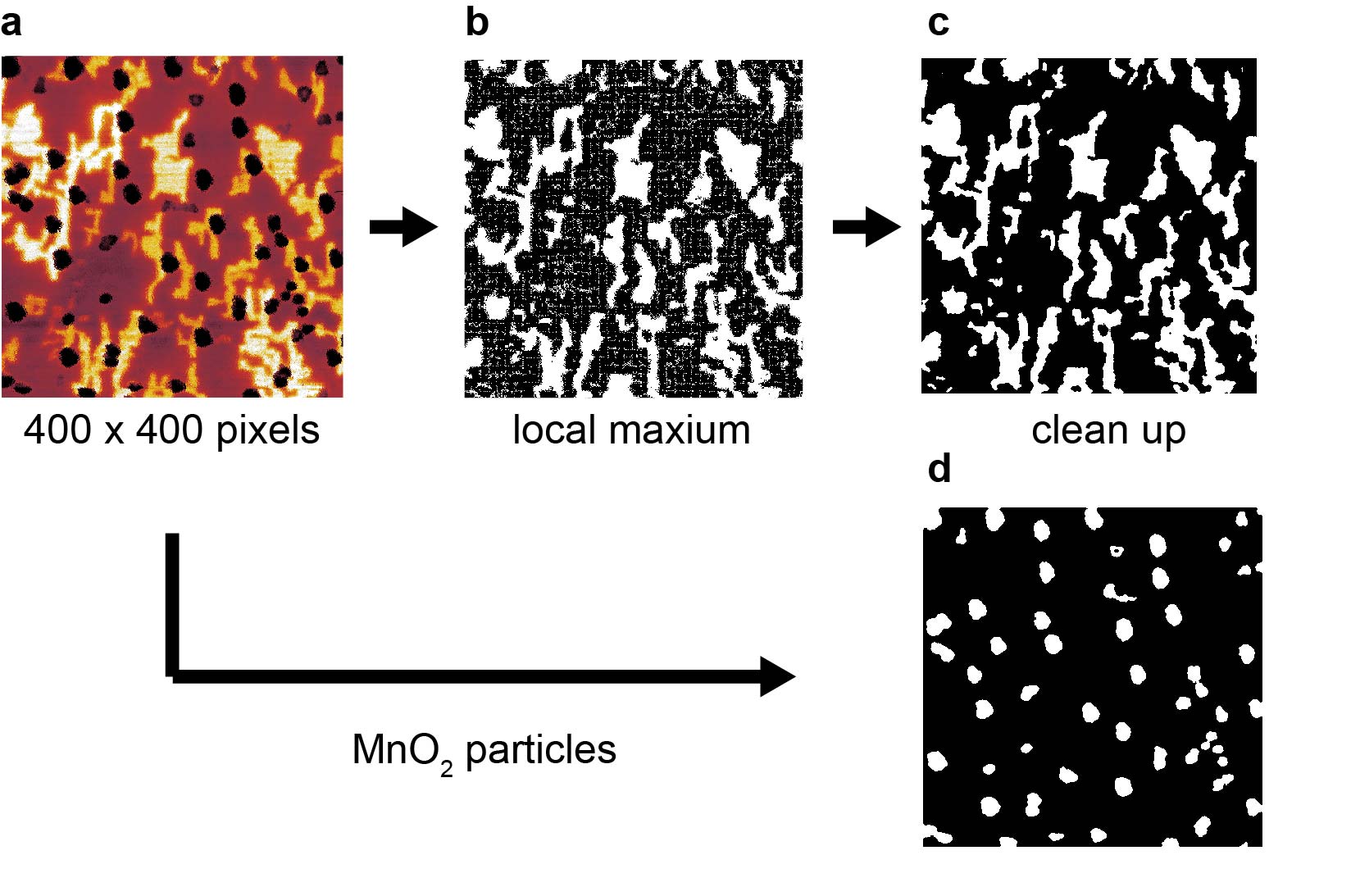}
	\caption{Method for extracting the FM-M fraction. }
	\label{sfig4}
\end{figure}


The areal fraction of the FM-M domains was obtained by the following procedure. 
\begin{enumerate}
	\item Divide the MIM image into 10 $\times$ 10 pixels images.
	\item Find a local maximum corresponding to the domains of each images.
	\item Put the images back together to obtain the image in \ref{sfig4}(b). 
	\item Domains smaller than 20 pixels in size were considered noise and were removed resulting in the image in \ref{sfig4}c. 
	\item Using the same procedure, we extract the area cover by Mn$_3$O$_4$ particles \ref{sfig4}(d), where states underneath are unknown. 
	\item The FM-M fraction was obtained by counting the number of the white pixels in c, and divided by number of the black pixels in \ref{sfig4}(d). 
\end{enumerate}

\begin{figure}[!]
	\centering
	\includegraphics[width=3in]{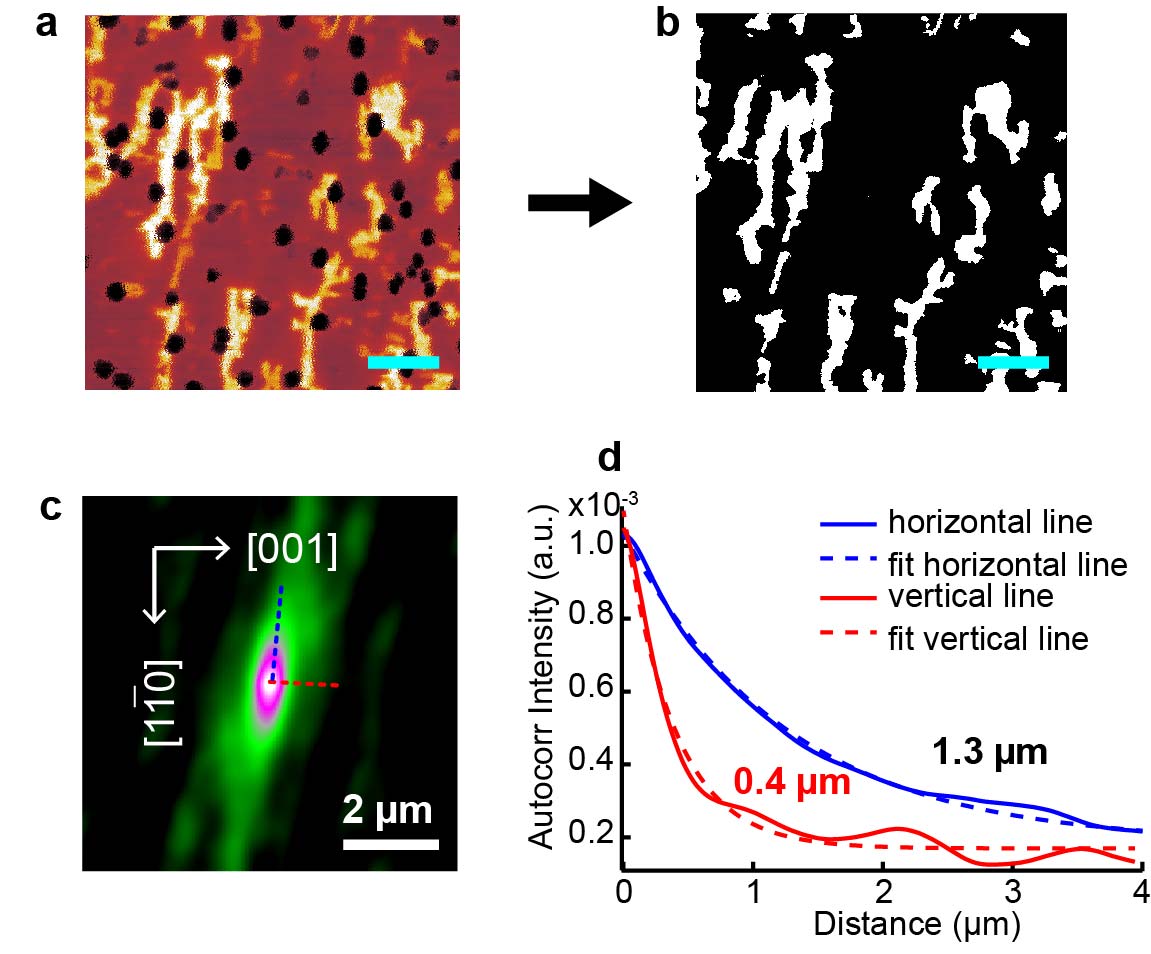}
	\caption{Method for autocorrelation analysis.}
	\label{sfig5}
\end{figure}

In order to show the preferential orientation of the FM-M domain, we performed autocorrelation analysis. The autocorrelation analysis presented in the main text was obtained with the following method. The metallic domains were extracted from FIG. \ref{sfig5}(a) (a MIM image at 12 K after a slow cool down) using a crude threshold value, resulting in the black-white image in FIG. \ref{sfig5} (b). Autocorrelation image is shown in FIG. \ref{sfig5} (c). FIG. \ref{sfig5} (d), Plot of the linecuts along the vertical (blue) and horizontal (red) lines of the autocorrelation. The characteristic length along the [1$\bar{1}$0] and [001] axes were obtained by fitting exponential functions to the vertical and horizontal linecuts. 

\begin{figure}
	\centering
		\includegraphics[width=3in]{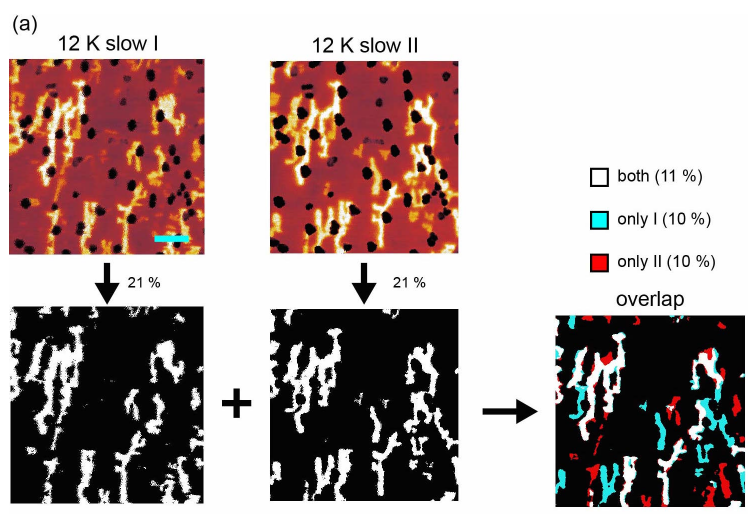}
		\caption{Overlap of the FM domains, and cross correlation coefficient at 12 K.}
			\label{sfig6}
\end{figure}

To illustrate the change of the FM-M domains before and after waiting, we overlaid two MIM images using the following procedure. Using FIG. \ref{sfig6}(a), MIM images at 12 K after the two slow cool downs, as an example, the images were first aligned by optimizing the standard cross-correlation coefficient of the raw images. We then extracted the black-and-white images showing only the FM-M domains before overlaying them on top of each other. White, blue and red colors are assigned to the region appeared in both images, only the first and only the second images accordingly. The overlaying method is also useful for observing the relaxation of the FM-M domains.

To further quantify the similarity of two MIM images, we calculated the cross correlation coefficient ($r_{xy}$). The $r_{xy}$ calculation also employed the black-and-white images to prevent artifact from the MIM signal variation in the actual images. The white region is assigned the value 1 and the black region the value -1. The $r_{xy}$ is computed from the sum of the dot product of these two images, divided by the number of pixels in the image.

\subsection{Relaxation time }

\begin{figure}
	\centering
		\includegraphics[width=3.2in]{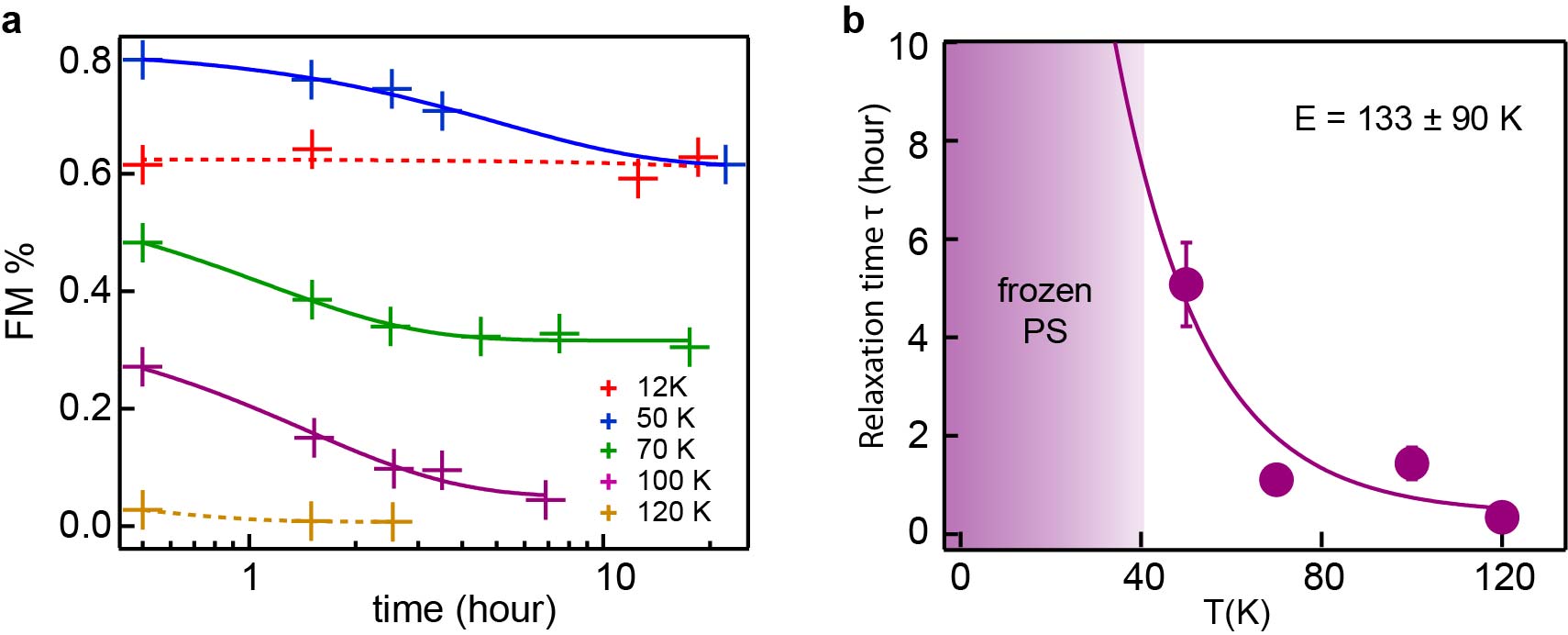}
		\caption{ Time dependence FM-M fraction and relaxation time.}
			\label{sfig7}
\end{figure}

FIG. \label{sfig7}(a) plots the FM-M fraction as a function of time (+) where the lines are exponential fits using (a) plots the FM-M fraction as a function of time (+) where the lines are exponential fits using 

\begin{equation}
M(t) = M_0 + Ae^{-(t-t0)/\tau}
\end{equation}

Although most transport studies ( for example PRB \textbf{64}, 104409 (2001)) prefer fitting magnetziation data with an logaritmic function, we choose a exponential functionin our fit. This is because the data points from our scans were taken at an hour apart, which is a coarse time scale compared to the transport data. The scanning data captures a slow relaxation process and is suitable for an exponential fit. The relxation time $\tau$ is plotted as a function of temperature in FIG. \label{sfig7}(b). The purple line is just a guide to the eyes. Fitting temperature dependent relaxation times with either the Vogel-Fulcher law or the power law causes large error due to few data points; and thus is omitted here.   

\newpage 

\subsection{Gradual Colorscale Images}

\begin{figure}
	\centering
		\includegraphics[width=5in]{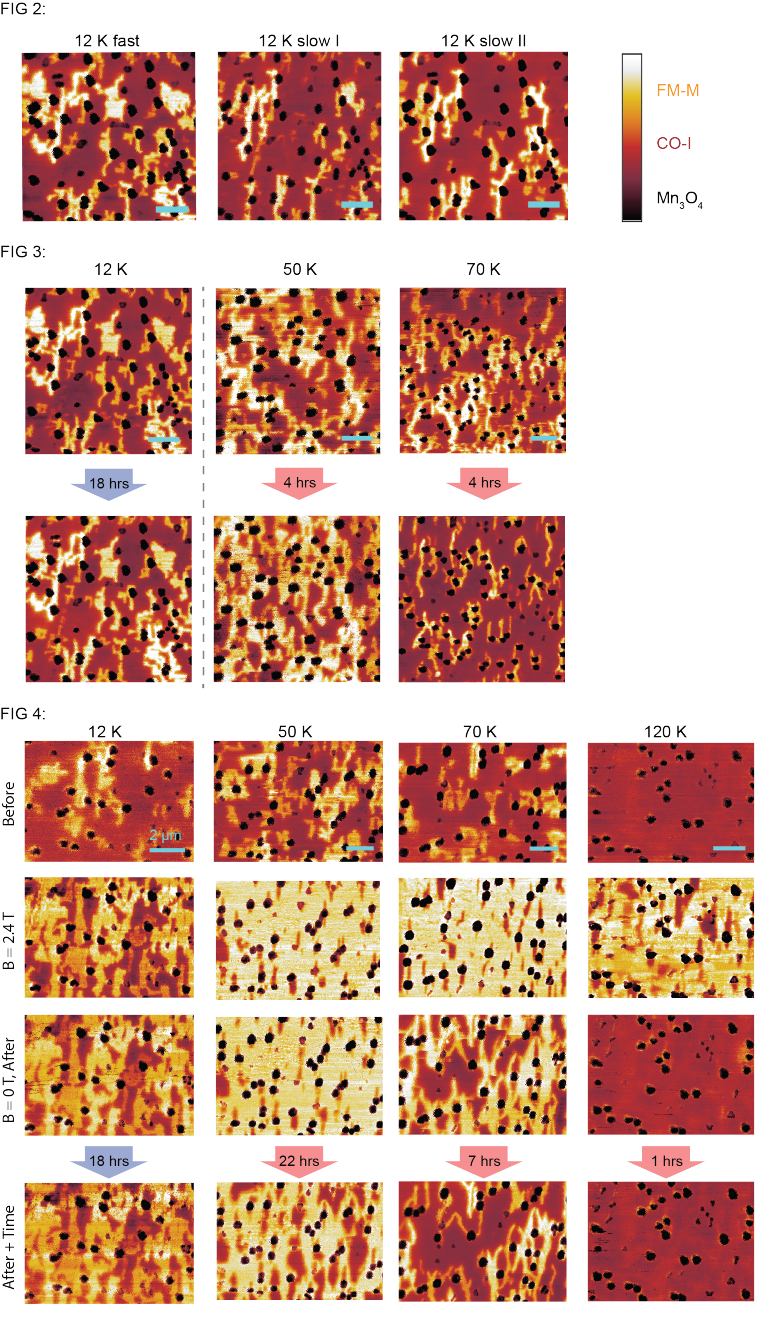}
		\caption{MIM images from the main text in a gray colorscale.}
			\label{sfig8}
\end{figure}

\end{document}